\documentclass[useAMS,usenatbib]{mn2e}

\usepackage{graphicx}
\usepackage{times}
\usepackage{natbib}

\newcommand{\apjs}{ApJS}

\newcommand{\apjl}{ApJ Letters}

\newcommand{\aap}{A\&A}

\begin{document}
 
\title[Do radio relics challenge DSA?]{Do radio relics challenge diffusive shock acceleration?}
\author[F. Vazza, M. Br\"{u}ggen]{F. Vazza$^{1,2}$, M. Br\"{u}ggen$^{1}$
\thanks{%
 E-mail: franco.vazza@hs.uni-hamburg.de}\\
 %EndAName
$^{1}$ Hamburger Sternwarte, Gojenbergsweg 112, 20535 Hamburg, Germany \\
$^{2}$ INAF/Istituto di Radioastronomia, via Gobetti 101, I-40129
Bologna, Italy}

\date{Accepted ???. Received ???; in original form ???}
\maketitle

\begin{abstract}

Radio relics in galaxy clusters are thought to be associated with powerful
shock waves that accelerate particles via diffusive shock acceleration (DSA). Among the particles accelerated by DSA, relativistic protons should outnumber electrons by a large factor.
While the relativistic electrons emit synchrotron emission detectable in the radio band, the protons interact
with the thermal gas to produce gamma-rays in hadronic interactions. Using simple models for the propagation of shock waves through clusters, the distribution of thermal gas and the efficiency of DSA, we find that the resulting hadronic $\gamma$-ray emission lies very close or above the upper limits from the FERMI data on nearby clusters. This suggests that the relative acceleration
efficiency of electrons and protons is at odds with predictions from DSA. The inclusion of re-accelerated  "fossil"  particles does not seem to solve the problem. Our study highlights a possible tension of the commonly assumed  scenario for the formation of radio relics and we discuss possible solutions to the problem.

\end{abstract}

\label{firstpage}

\section{Introduction}
\label{sec:intro}

Giant radio relics are steep-spectrum radio sources that are found in the 
outer parts of galaxy clusters, $\sim 0.5-3$ Mpc from their centres \citep[e.g.][and references therein]{br11,fe12}. 
They are almost exclusively found in perturbed clusters, and there is good evidence for their
 association with powerful shock waves triggered by mergers \citep[][]{1998A&A...332..395E,hb07}. 
%Among them are double-relics with the two large-scale emission at different sides of the host cluster center \citep[e.g., ][]{2012MNRAS.426...40B}, which likely track the expansion of  internal merger shocks \citep[e.g.][and references therein]{vw12sim}.\\
Despite its low efficiency at low Mach numbers, diffusive shock acceleration (DSA, \citealt[e.g.][for modern reviews]{2012JCAP...07..038C,kr13}) has been singled out as the most likely mechanism to produce the relativistic electrons that power radio relics. According to theory, only a tiny fraction ($\ll 0.01$) of the kinetic power of a shock is necessary to 
accelerate electrons, and the observed power-law spectra ($s \sim 1$ where $I(\nu) \sim \nu^{-s}$ is the radio spectrum) are naturally produced by DSA \citep[e.g.][]{hb07}.\\
Cosmological simulations have been performed to investigate the complex ICM dynamics that may produce relics \citep[e.g.][]{pf07,ho08,sk13}. However, it is found that the acceleration of relativistic protons by the same shocks must also cause a significant level of hadronic $\gamma$-ray emission \citep[][]{1980ApJ...239L..93D,bbp97,bl99, pf07,alek12,va13feedback,fermi13}.\\
In this work, we propose that a combination of radio observations that reveal the relativistic electrons and $\gamma$-ray observations that can constrain the density of relativistic protons provides a good test for DSA at radio relics. We resort to a semi-analytical approach to estimate the total CR-energy injected by shocks over the course of their existence inside the cluster. Our results show that for several clusters with relics and FERMI data, the combination of Mach number, magnetic field strength and gas density that is required to explain the radio relics produces $\gamma$-ray emission above the upper limits.
This suggests a possible tension between the commonly assumed acceleration efficiencies in the Mach number range $2 \leq M \leq 5$.  Although the present-day $\gamma$-ray observations are not deep enough to exclude DSA for the entire set of our clusters, this analysis favours a large acceleration efficiency ($\xi_{\rm e} \geq 10^{-4}-10^{-3}$) at weak shocks, and a larger energy ratio between injected relativistic electrons and protons than is commonly adopted ($K_{\rm e/p} \geq 0.1$), at odds with the standard view of shock acceleration that works seemingly well in supernova remnants.

\section{The sample}
We build a sample of radio relics by combining the radio data in the catalog by \citet{fe12} and the $\gamma$-ray data in the energy range [0.2-100] $\rm GeV$ by the FERMI collaboration
\citep[][]{ack10,fermi13}. This yields 6 objects: Abell 2163, Abell 2256, Abell 2744, Abell 754, Abell 3376 and Abell 1656 (COMA), for which the main parameters used in our calculation are shown in Table 1.\\
In addition to these objects, we include 13 {\it double} relics from the catalog by \citet{2012MNRAS.426...40B}, which are not included in the set of published FERMI data by \citep[][]{ack10,fermi13}. We included only double radio relics since they can most confidently be traced back to merger shocks. Where spectral indices were not known (A3365, A3367), we assumed a spectral index according to the observed relation between the projected largest linear size of relics, $LLS$, and their integrated radio spectrum, $s \sim 2 + \log_{\rm 10}(LLS/\rm 300 kpc)$, as in \citet{vw09}.
In those clusters, where the gas temperature could not be determined from X-ray observations, we assumed a  post-shock temperature based on the $L_{\rm X} - T_{\rm 500}$ relation \citep[][]{2009A&A...498..361P}.\\
In order to compare to the prediction of hadronic emission from this second set of relics, we used recent upper limits from a collection of stacked Fermi-LAT count maps, recently performed by \citet{2013arXiv1308.6278H} and \citet{2012A&A...547A.102H}, which are the deepest up to date.  The result of the stacking cannot be strictly used on a single-object comparison, since the properties of the single object might differ significantly from the properties
of the clusters in the stacked sample. In the following, the comparison of our forecasts of hadronic emission  with the stacking upper-limits will rely on the assumption that the
population of non cool-core clusters (for which the limits of \citealt{2013arXiv1308.6278H} have been derived) and the one of clusters hosting double relics are similar and affected by a similar scatter.

\begin{table*}
\label{tab:tab1}
\caption{Parameters for the radio relics and clusters considered in this paper. The data are taken from \citet{fe12} (radio power, spectral index, size and cluster-centric distance of the relics) and \citet{ack10,fermi13} ($\gamma$-ray emission, $L_{\rm X}$ and $T_{\rm 500}$). The last four columns show the values assumed for the downstream/upstream gas density, for the Mach number and for the electron acceleration efficiency necessary to match the observed radio power, in the model discussed in Sec.\ref{sec:radio}. In the case of A3376 there is some disagreement between the radio \citep[][]{2012MNRAS.426.1204K} and X-ray derived Mach numbers \citep[][]{2012PASJ...64...67A}. The $\gamma$-ray emission of A2256, A754, A3376 and COMA have been derived from \citet{fermi13} after correcting for the slightly different energy band.}
\centering \tabcolsep 5pt 
\begin{tabular}{c|c|c|c|c|c|c|c|c|c|c|c}
   object & z & $\rm log P_{\rm 1.4GHz}$  & $R_{\rm relic}$ &  LLS  &  $L_{\rm X}$  &   T& $N_{\gamma}$ 
[0.2-100] & $n_{\rm d}$& $n_{\rm u}$ & $M$ & $\xi_{\rm e}$ \\
 
            &          &      [W/Hz] & [Mpc]  & [Mpc] & [$10^{44}$  erg/s] & [keV] & [${\rm ph/(s 
\, cm^2)}$] & [1$\rm /cm^3$] & [1$\rm /cm^3$] &  & \\ \hline% & [$\rm \mu G$] \\ \hline
A2163 & 0.2030  & 24.32  & 1.17  & 0.48  &   22.73&   13.3 & $5.51 \cdot 10^{-9}$ & $6.5 \cdot 10^{-4}$ & $2.0 \cdot 10^{-4}$ & 5.5 & $2.41 \cdot 10^{-5}$\\%& 0.85  \\
A2256  &0.0581    & 24.56  & 0.44  & 1.13  &   3.75   &   6.6 & $2.76 \cdot 10^{-10}$ &  $2.7  \cdot 10^{-3}$ & $9.1  \cdot 10^{-4}$ & 3.3 & $2.65 \cdot 10^{-6}$ \\% & 7.0 \\
A2744 & 0.3080   &  24.71  & 1.56  & 1.62   & 12.86   &      10.1  & $2.49 \cdot 10^{-9}$ &  $3.1 \cdot  10^{-4}$& $1.0  \cdot 10^{-4} $ & 4.5 &$6.93 \cdot 10^{-5}$  \\% & 0.95 \\
A754 & 0.0542     & 23.67     & 0.5 & 0.8 & 2.21 & 9  & $1.05 \cdot 10^{-9}$&  $1.9  \cdot 10^{-3}$ & $8.6  \cdot 10^{-4}$ & 2.2 & $1.61 \cdot 10^{-6}$ \\% & 1.8\\
A3376W & 0.0468   & 23.88   & 1.43 & 0.8 &  1.08 & 4.3 &  $6.26\cdot 10^{-10}$ &  $1.7  \cdot 10^{-4}$ & $4.9  \cdot 10^{-5}$ & 3.3 & $6.12 \cdot 10^{-5}$  \\% & 3.0 \\
A3376E & 0.0468   & 23.79   & 0.52 & 0.95 &  1.08 & 4.3 &  $6.26 \cdot 10^{-10}$&  $1.25  \cdot 10^{-3}$ & $4.2  \cdot 10^{-4}$ & 2.5 & $4.45 \cdot 10^{-6}$\\% & 3.0 \\
A1656 & 0.0231   & 23.49   & 2.2   & 0.85 & 3.99 & 8.3 & $6.62\cdot 10^{-10}$ & $1.0  \cdot 10^{-4}$ & $3.6  \cdot 10^{-5}$ & 3.3 & $1.63 \cdot 10^{-5}$\\% & 2.0\\
\end{tabular}
\end{table*}

\section{Modelling of radio relics}
\label{sec:radio}

Throughout the paper, we assume that all relics trace outward propagating shocks that have swept through the cluster volume{\footnote {At least in the case of the Coma cluster, A1656, this assumption is questionable because the shock may be an infall shock \citep[][]{2013MNRAS.433.1701O}.}}.
In order to estimate the radio emission by relativistic electrons, we use the analytic model developed by \citet{hb07}, hereafter HB07.
This model assumes an exponential cutoff in the energy distribution of electrons, determined by the balance of the acceleration rate and of the (synchrotron and Inverse Compton) cooling rate.
In the downstream region, DSA is assumed to generate supra-thermal electrons
that follow a power-law in energy. While the gas is advected with the
downstream plasma, the relativistic electrons cool because of synchrotron and inverse Compton losses. The maximum energy of the electrons is a decreasing function of the distance from the shock front, and the radio emission diminishes accordingly. The total emission in the downstream region is finally obtained by summing up all contributions from the plasma from the shock front to the distance where the electron spectrum is too cool to allow any further radio emission at observational frequencies.
The monochromatic radio power at frequency $\nu$, $P_{\nu}$, depends on the shock surface area, $S$, the downstream electron density, $n_{\rm d}$, the electron acceleration efficiency, $\xi_{\rm e}${\footnote{We note that, in agreement with most of the literature on DSA of electrons, we include in our definition of the electron acceleration efficiency the dependence on $M$ in Eq.\ref{eq}, i.e. $\xi_{\rm e} = \xi_{\rm e,0} \Psi(M)$, where $\xi_{\rm e,0}=0.05$ and $\Psi(M)$ is given in Eq. (31) of HB07.}} , the downstream electron temperature,  $T_{\rm d}$,  the spectral index of the radio emission, $s$, and the relic magnetic field, $B$, as in Eq.32 of HB07:
\begin{equation}
P_{\nu}\propto S\, \cdot n_{\rm d} \cdot \xi_{\rm e} \cdot \nu^{-\delta/2} \cdot T_{\rm d}^{3/2} \frac{B^{1+\delta/2}}{(B_{\rm CMB}^2+B^2)}
\label{eq}
\end{equation}
The shock surface is approximated as the square of the largest linear size of the relic. We fix the up-stream gas density at the relic using a $\beta$-model profile for each host cluster, with $\beta=0.75$ and the core radii scaling as $r_{\rm c}=r_{\rm c, Coma}(T_{\rm d, Coma}/T_{\rm d})^{1/2}$ (which follows from the self-similar scaling), where $r_{\rm c, Coma}=290$ kpc. Then, we fix the Mach number $M$ from the spectral index of the radio spectrum at the injection region via $M= \sqrt \frac{\delta+2}{\delta-2}$ ($\delta$ is the slope of the particle energy spectra and $\delta=2 s$).
From the Mach number, $M$, we compute the downstream density at the relic location using Rankine-Hugoniot jump conditions, using a ratio of specific heats $\gamma=5/3$. Similarly, for the temperature we assume $T_{\rm 500}$ for each cluster as the post-shock value of temperature and we derive the pre-shock temperature again using shock jump conditions.\\
Now we can calculate the kinetic power dissipated at the shock,
$\Phi_{\rm kin} \propto n_{\rm u} V_{\rm s}^3  S/2 $, where $V_{\rm s}=M c_{\rm s}$ ($c_s \propto \sqrt{T_{\rm u}}$). Next we can compute the synchrotron power of each relic, provided we know (i) the magnetic field in the relic, $B_{\rm d}$ and (ii) the electron acceleration efficiency, $\xi_{\rm e}$. We explore two possible ways of 
proceeding. \\
In the first case, we fix the magnetic field and derive the acceleration efficiency of electrons by equating $P_{\nu}$ to the observed radio power.
We then set the CR-proton acceleration efficiency to $\eta=\xi_{\rm e}/K_{\rm e/p}$, where $K_{\rm e/p}$ is the electron-to-proton ratio. The power in CRs is then given by $\Phi_{\rm CR}=\eta \Phi_{\rm kin}$.
The estimate of  $K_{\rm e/p}$ is highly uncertain and presently cannot be derived from first principles. A value of $K_{\rm e/p} \sim 0.01$ is commonly assumed for the Milky Way, since about $\sim 1 $ percent of the observed Galactic flux of CRs with $E \approx \rm GeV $ is due to electrons \citep[e.g.][]{2002cra..book.....S}. On the other hand, the modelling of the spectra of young supernova remnants 
generally requires $K_{\rm e/p} \sim 10^{-3}-10^{-5}$ to fit the observational data \citep[e.g.][]{2009A&A...505..169B,2009MNRAS.392..240M,2010ApJ...712..287E,2012ApJ...746...82F}. In our computation here we set the electron to proton ratio  to $K_{\rm e/p}=0.01$, which is the most conservative assumption since it already minimizes the acceleration of CR protons for a given amount of CR electrons. As we will see in the next sections, already in this case we tend to predict too large values of CR proton acceleration in several systems and therefore the adoption of lower (and perhaps more realistic) values of $K_{\rm e/p}$ only exacerbates the problem.\\
In a second model, we will assume that $\eta$ depends on $M$ according to the results of \citet{kr13}, who studied the  dissipation of energy into CRs as a function of the sonic Mach number and also included the magnetic 
field amplification by streaming instabilities excited by CRs \citep[see][for more details]{kr13}. The additional reaccelerating of pre-existing CRs can be included by using modified versions of $\eta(M)$ that depend on the ratio of the non-thermal to thermal energy densities, $\epsilon$, in the upstream region. To this end we use a fit to their Figure 3. The electron acceleration efficiency is given by $\xi_{\rm e}(M) = \eta(M) K_{\rm e/p}$. Now, $B_{\rm d}$ is a free parameter, which we fix by matching the observed radio power given by Eq.\ref{eq}, where all the other parameters have been fixed as before and as given in Table 1. However, unlike in the previous model we fix the Mach number at the current relics to $M=2$, which is a conservatively low value. This then yields the corresponding $\eta(M)$ given a value of $\epsilon$. In this study we tested $\epsilon=0.01$ and $\epsilon=0.05$, i.e. in the range of what found in our previous cosmological simulations \citep[][]{scienzo} and of the order of the maximum allowed budget of CRs by the most recent upper limits \citep[][]{fermi13} .
Because of their large life-time ($\geq 10$ $\rm Gyr$ ), the pre-existing CR-protons could come from previous internal and external shocks. In contrast, the pre-existing CR-electrons must be produced locally, e.g. by turbulent re-acceleration or hadronic collisions, because of their short  radiative life time ($<10^8$ $\rm yr$ for a $B \sim \mu G$ magnetic field). As in \citet{kr13}, we assume that the blend of several populations of pre-existing CR-electrons are characterised by an energy spectrum  $\delta_e = (4\delta+1)/3$, where $\delta$ is the energy spectrum derived at the relic as before.  This follows from the fact that the general effect of shock re-acceleration is that of
generating particle spectra that are flatter than the spectrum of the seed particles if the re-acceleration Mach number is strong, while conserving the slope of the seed energy distribution if the re-acceleration Mach number is weaker. In both cases, shock re-acceleration increases the normalization of the distribution more than what simple adiabatic compression does \citep[][]{2004NuPhS.136..208B,2005ApJ...627..733M}.
In order to predict the gamma-ray emission throughout the cluster, we must make some assumptions about the properties of the shock as they travel thorugh the cluster. We assume that the shock surface scales with cluster-centric distance, $r$, as $S(r)=S_0 (r/R_{\rm relic})^2$ (i.e. the lateral extent of the shock surface is set to largest linear size of the relic, which decreases with $\propto r$ inwards). The total energy dissipated in CRs is given by $E_{\rm CR}= \int_{r_c}^{R_{\rm relic}}{\eta \Phi_{\rm kin}(r) dt }$, where $dt=dr/V_{\rm s}$ and $r_c$ is the cluster core radius. The shock surface and strength will vary with radius, and so will $\eta(M)$. For the scaling of $M$ with $r$ we inverted the relation between the radio spectral index and the injection
Mach number, using the results of \citet{vw09}, who observed flatter spectral indices of 
radio relics increasing with the distance from the cluster centre and approximate it as $M=M_0 (r/R_{\rm relic})^{1/2}$. Although this observed relation has large uncertainties, such gentle functional dependence with radius is in line with cosmological simulations \citep[e.g.][, Fig.15]{va10kp}.\\
The lower integration limit in the equation for $E_{\rm CR}$ is the core radius, $r_c$, meaning that the shocks are assumed to be launched only outside of the cluster core, which is also supported
by cosmological numerical simulations \citep[][]{va12relic,sk13}. 
%Since the density along the merger axis of clusters is enhanced significantly, the gas density along the swept-up volume is increased by a factor of $\delta_{\rm \rho}=2$ with respect to the $\beta$-model.% 
Finally, we neglect the presence of gas clumping and of the enhanced gas density along the merger axis, which would only increase the hadronic emission, since $\epsilon_\gamma \propto n^2$.\\
We compute the hadronic $\gamma$-ray emission from the CR-protons as in \citet{pe04} and \citet{donn10}, and assume that the momentum spectrum of protons follows a power law with a slope $\delta_{\gamma} \approx \delta$, and a threshold proton energy $E_{\rm min}=780 ~\rm MeV$. The spectral index is fixed to the radio spectral
index \citep[e.g.][]{ka12}. Unlike in \citet{pe04} and \citealt{donn10}, we use  the parametrisation of the proton-proton cross section given by \citet{2006PhRvD..74c4018K}.
%, which only leads to small differences in our energy range. 
\begin{figure}
    \includegraphics[width=0.45\textwidth]{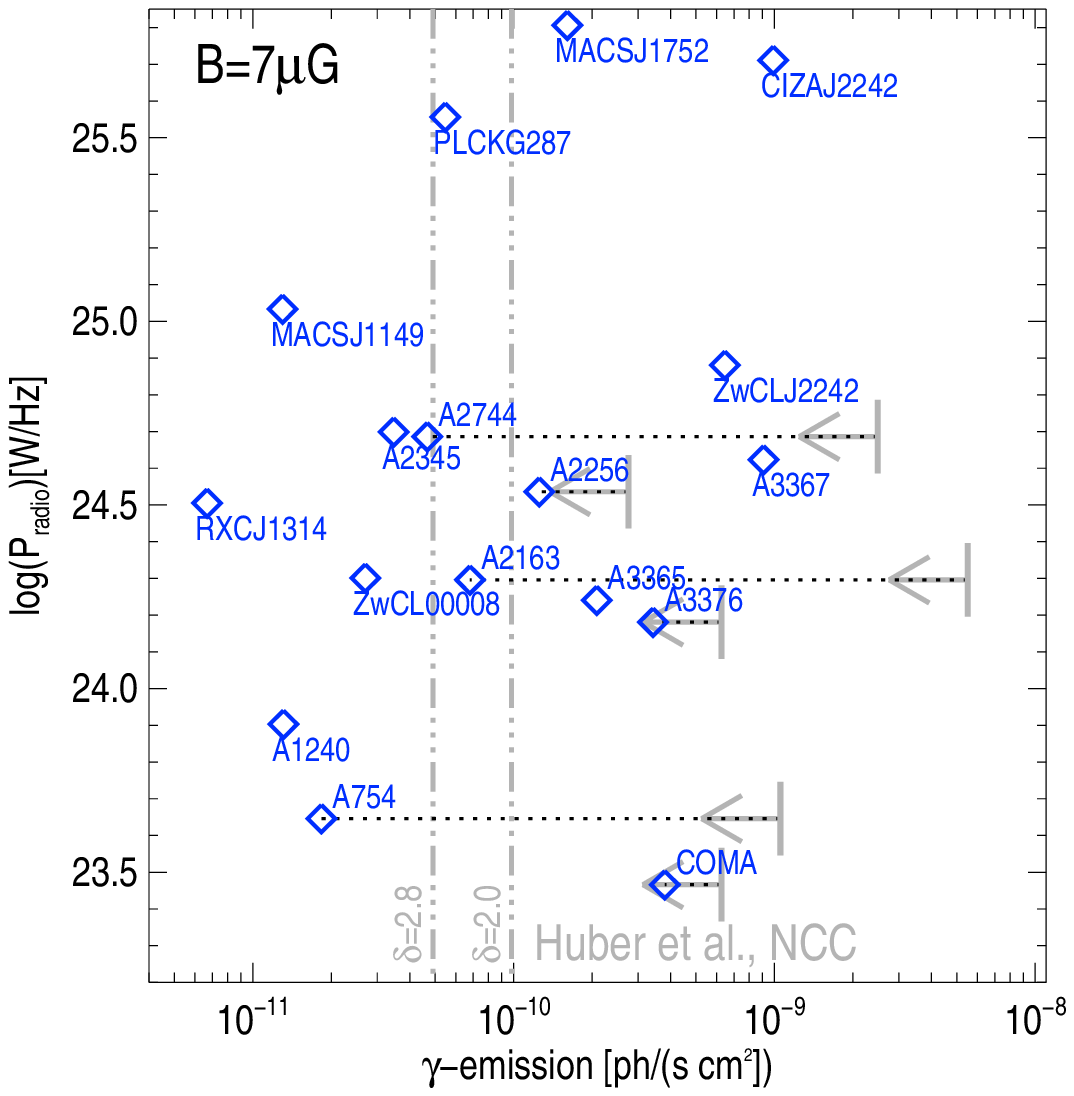}
    \includegraphics[width=0.45\textwidth]{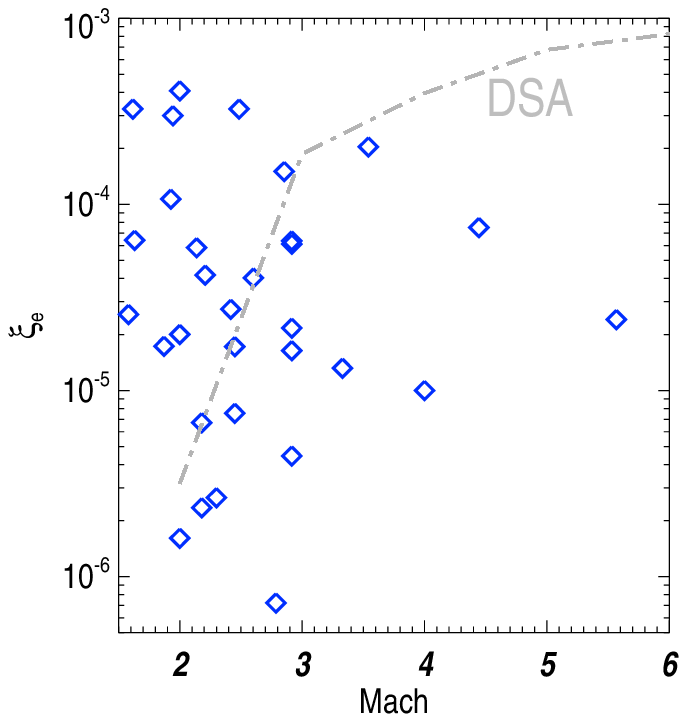}
         \caption{Top panel: synchrotron power at $1.4$ GHz and estimated 
hadronic $\gamma$-ray fluxes for our sample (blue squares). We fix $B_{\rm d}=7 \mu G$ and we constrain $\xi_{\rm e}$ by matching the observed radio power. The gray arrows show the FERMI upper limits for single sources \citep[][]{ack10}, while the vertical gray line shows the result of the stacking by Huber et al. (2013) for $\delta=2.0$ or $\delta=2.8$. The bottom panel
shows the values of $\xi_{\rm e}$ for each single relic of the sample. The additional grey line shows the expectation from DSA, as in in Fig.3 of \citet{kr13}.}
  \label{fig:first}
\end{figure}

\section{Results}

\subsection{Fixed magnetic field}
\label{subsec:first}

In our first model, we fix the magnetic field for each relic and derive the values of $\xi_{\rm e}$ and $\eta=\xi_{\rm e}/K_{\rm e/p}$ that are necessary to match the observed radio power. To be conservative, we choose $B_{\rm d}=7 \mu G$, which is the largest values of the magnetic field inferred for the radio relic in CIZA J2242.8+5301 \citep[][]{vw10}. By doing so, we are minimising the electron energy required to match the observed radio power, and hence the CR-proton energy used to compute the hadronic emission from the host clusters.\\
The first panel in Figure \ref{fig:first} shows the result of this test for our sample, and compare it to the $\gamma$-ray limits from \citet{ack10,fermi13} and Huber et al. (2013), within the energy range [0.2-100] $\rm GeV$. The FERMI data cover the cluster volume inside $R_{\rm 500}$ of clusters, which is about one order of magnitude larger than the volume
swept up by our shocks, and include the core region of clusters which we exclude. Hence, any additional source of CRs within the ICM would enhance the total hadronic emission predicted by our model. In the double relic systems, the radio and the $\gamma$-ray fluxes are added and the swept-up volume is doubled.\\
No $\gamma$-ray fluxes above the FERMI limits are predicted for our relics, if we compare with
the single object data provided by \citet{ack10} and \citet{fermi13}. However, for about one half of the sample we predict levels of hadronic emission that are larger than the upper limits of the
stacking analysis by Huber et al. (2013), considering a range of possible proton energy spectra ($\delta=2.0-2.8$). These limits have been derived for a sample of 32 non-cool-core (NCC) clusters, and they cannot be used on a single-object basis. Should clusters with radio relics have an intrinsically larger $\gamma$-ray emission than NCC clusters, the stacking analysis of the first population 
would lead to a higher upper limit. At this level we can only rely on the assumption that the two populations are similar in their content of CRs, which is reasonable following the scenario in which cluster mergers are both responsible for the release of 
non-thermal energy and for the disappearance of cool cores within clusters. With this caveat, we suggest that the hint of a tension is shown by this test, since many clusters hosting double relics have a predicted $\gamma$-ray emission $\sim 5-10$ times above the available upper-limits from the stacking of the NCCs.\\
Another hint for tension with DSA results is shown in the second panel in Fig.\ref{fig:first}, where the values of $\xi_{\rm e}$ needed to match the observed radio power at $1.4$ GHz are shown assuming $B_{\rm d}=7 \mu G$. 
In the same figure we show the predicted dependence of $\xi_{\rm e}$ on Mach numbers according to the numerical study by \citet{kr10}. For about half of the relics of our sample the required value of $\xi_{\rm e}$ is large compared to the theoretical expectation of DSA for weak shocks, $M \leq 3$. The tension can be even larger if magnetic fields are lower than the high $7 \mu G$ considered here.\\
Our result for weak shocks confirms the result of recent studies that highlighted the importance of shock re-acceleration of CRs in weak shock regime, in order to explain observations of several relics \citep[][]{ka12,2013MNRAS.435.1061P}.  However, rather high values of $\xi_{\rm e}$ even for stronger shocks seem necessary for $M \geq 3$ relics, add odds with the expectation of DSA theory.
In the next section, we investigate how the inclusion of shock re-acceleration can change this picture.

\begin{figure}
    \includegraphics[width=0.45\textwidth]{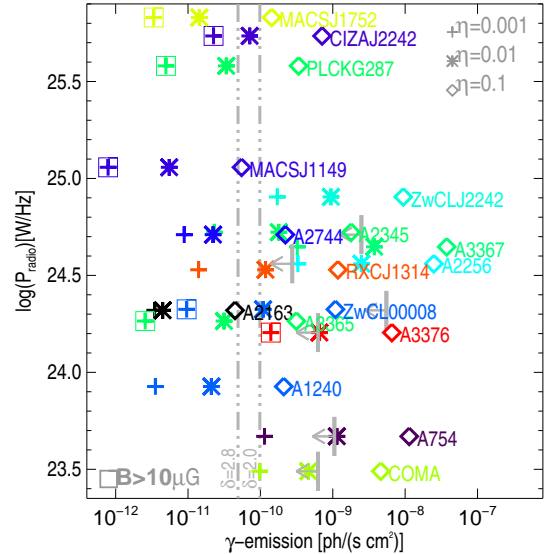}
    \caption{Synchrotron power at $1.4$ GHz and estimated 
hadronic $\gamma$-ray emission obtained by imposing three different fixed acceleration efficiencies for CR protons, 
The magnetic field is derived by matching the observed radio power of real relics, and with an additional big square mark the relics for which the solution
yields $B \geq 10 \mu G$. The $\gamma$-ray limits are as in Fig.\ref{fig:first}.}
  \label{fig:new}
\end{figure}

\begin{figure}
    \includegraphics[width=0.45\textwidth]{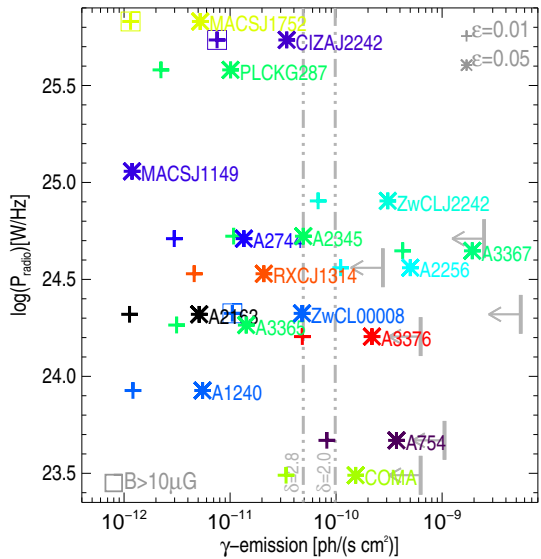}
    \caption{Synchrotron power at $1.4$ GHz and estimated 
hadronic $\gamma$-ray emission for three models of DSA with an increasing level of re-accelerated
CRs (Sec.\ref{subsec:second}). The magnetic field is derived by matching the observed radio power of real relics, and with an additional big square mark the relics for which the solution
yields $B \geq 10 \mu G$. The $\gamma$-ray limits are as in Fig.\ref{fig:first}.}
  \label{fig:second}
\end{figure}

\subsection{Variable magnetic field and shock reacceleration}
\label{subsec:second}

In our second model, we parametrise the acceleration efficiency of CRs assuming a functional dependence on the Mach number, $\eta(M)$, predicted by DSA following \citet{kr13} (or using a fixed value), and we constrain the magnetic field by matching the observed synchrotron emission.\\
As a first simple case, we impose a constant value for all Mach numbers and explored the cases $\eta=0.1$, $10^{-2}$ and $10^{-3}$, see Fig.\ref{fig:new}.  If a low acceleration efficiency is assumed, the predicted hadronic $\gamma$-emission is below the FERMI limits in most cases, but in half of the objects the required magnetic field becomes unrealistically large $\gg 10 \mu G$ (up to $\sim 10^{3} \mu G$ in a few cases), at odds with observations \citep[][]{fe12}{\footnote{In principle, CR-driven amplification can yield very large values of magnetic field ($\sim 10^2 \mu G$) in the precursor of strong supernova shocks \citep[e.g.][]{2012JCAP...07..038C,2013MNRAS.431..415B}. However, in the case of galaxy clusters such large values have never been observed and, furthermore, recent theoretical work suggests that the CR-driven amplification in weak merger shocks can produce at most $\sim 5 \mu G$ in the ICM \citep[][]{2013MNRAS.tmp.2295B}.}} On the other hand, if we assume a higher acceleration efficiency the synchrotron emission
at relics is generally recovered using a value of the magnetic field consistent with observations, yet the hadronic emission
from the downstream exceeds the FERMI limits.\\
For weak shocks ($M \leq 3-4$) the re-acceleration of ``fossil'' electrons \citep[][]{2005ApJ...627..733M,ka12,2013MNRAS.435.1061P}, could alleviate the problem of the large acceleration efficiencies required in Sec.4.1. Fig.~\ref{fig:second} shows our results in this case. In this case a single injection model using
$\eta(M)$ from \citet{kr13} and without pre-existing CRs does not violate the constraints 
from FERMI for the hadronic emission (for the sake of clarity we omit these points from the Figure). However, again in more than 80 percent of cases the required magnetic field becomes way too large compared with observations. This supports the results of the previous section, and confirms similar conclusions by \citet{ka12} for their modelling of the relics in ZwCl0008.8+5215.
However, the models including re-acceleration generally yield much more reasonable values of magnetic fields at the cost of a higher energy dissipation into CRs.
The model with an energy ratio $\epsilon=0.01$ between seed CR-protons and thermal gas requires unrealistically
large values of $B$ in a couple of systems, while in all other cases the values are reasonable, $0.3 \mu G \leq B \leq 5  \mu G$. However, the re-acceleration models predict levels of hadronic $\gamma$-ray emission that are too large for many objects: $\sim 1/2$ of objects in the $\epsilon=0.05$ case
and $\sim 1/3$ of objects in the $\epsilon=0.01$ case have $\gamma$-ray emission above the stacking
upper limits by Huber et al. (2013). In some cases, the  $\epsilon=0.05$ model also predicts $\gamma$-ray
emission above or very close to the single object limits by FERMI 
(A3667, A2256, A3376, A754 and COMA, i.e. among the nearest galaxy clusters of the sample). 

%WHY?
%
%Larger values of $\epsilon$ The trend of the simulated data with the 
%increase of $\epsilon$ clearly shows that any larger value of $\epsilon$ cannot but make the comparison with FERMI limits worse. 
%

\section{Conclusions}
\label{sec:conclusions}

In this study we modelled a sample of observed radio relics using DSA and computed the resulting acceleration of CR-protons. We point out several issues with the standard explanation for radio relics: First (Sec. \ref{subsec:first}), if the magnetic field in the relics is fixed to a large value ($B_{\rm d}= 7 \mu G$) the observed radio power demands a high acceleration efficiency for electrons ($\xi_{\rm e} \sim 10^{-4}-10^{-3}$). In several cases, i.e. $\sim 10-10^2$ higher than what is typically attained in supernova remnant shocks \citep[e.g.][]{2012JCAP...07..038C}. As a result, the CR-proton acceleration efficiency is also large, and for a significant fraction of objects this yields hadronic $\gamma$-ray fluxes above the upper limits from the stacking of NCC cluster by FERMI. A complementary approach (Sec.\ref{subsec:second}) of assuming a functional dependence for the acceleration of CRs \citep[as in][]{kr13} and constraining the required value of $B_{\rm d}$ at relics yields other problems: either the magnetic field
values are unrealistically large ($\gg 10 \mu G$), as in the direct injection model, or the predicted hadronic $\gamma$-ray emission violates the FERMI upper-limits for several objects. \\
Relaxing any of the assumptions made in our modelling (e.g. absence of clumping, only moderate enhancement of the cluster profile along the merger direction, absence of any other source of CRs but the shock, neglect of the adiabatic compression of CRs, exclusion of the core regions for the launch of shocks, etc)  would increase, in some cases drastically, the predicted hadronic emission and thus
exacerbate the tension with the standard prediction of DSA. \\
Outside of cluster cores, the CR-energy is only marginally subject to hadronic and Coulomb losses, and CRs do not  move from their radius of injection, which is reasonable if the cosmic rays are frozen into the gas (owing to the tangled distribution of magnetic fields in the ICM,  which makes the diffusion time long (i.e. $\tau \sim 2 \cdot 10^8 \rm yr (R/Mpc)^2 (E/GeV)^{-1/3}$ for a constant $B=1 \mu G$ magnetic field and assuming Bohm diffusion, e.g. \citealt{bbp97}). In the case that CRs can sink further into the cluster atmosphere, they would get adiabatically compressed by the increasing ambient pressure. This effect would increase the CR energy density in the innermost regions and was not included in our simulations. We offer a few possible solutions to this problem:\\

(i) ``Geometrical`` arguments:  the shocks are launched only in the outer parts, very close to where the relics are presently observed. This is not in accord with what most simulations predict, even though there are cases where the radio relic appears to be very close to the cluster centre (e.g., A2256). 

(ii) The Mach numbers inferred from the radio spectrum are systematically overestimated. This would reduce the production of $\gamma$-ray emission in some systems (A3376, A2744), but require a much higher value for $B$. If particles are accelerated directly from the thermal pool and the assumptions of the test-particle regime hold, then the radio-derived Mach number
should be consistent with the Mach number derived from X-ray data. In fact, this is the case for all the other merger shocks confirmed at radio relics, except for the relic in 1RXS J0603.3+4214 \citep[][]{2013MNRAS.433..812O}. In addition, this cannot solve the problems for objects that are already believed to host weak shocks, such as A2256, A754 and Coma.

(iii) The ICM is filled with fossil CR electrons, but not with fossil CR-protons. In this case, shocks can energise the CR electron population without producing
hadronic $\gamma$-ray emission above the constraints from FERMI. It is difficult to imagine such a scenario since most acceleration mechanisms in the ICM accelerate CR-protons at least as much as electrons. 

(iv) The spectra of pre-existing CR electrons are significantly different from a power-law increasing at low energies and the bulk of the population is already highly relativistic ($\gamma \sim 10^3$). In this case, it is conceivable that  weak shocks can efficiently energise the electrons, without pushing the CR proton budget above the levels implied by FERMI. Again, it is not trivial to find circumstances under which the relativistic electrons are protected against expedient radiative losses.

(v) In principle, fast streaming of CRs can progressively deplete the downstream region of the shock and reduce the 
amount of hadronic emission \citep[e.g.][]{2011A&A...527A..99E}. However, the validity of this scenario is controversial \citep[e.g.][]{2013arXiv1306.3101D,2013MNRAS.434.2209W} and this mechanism has been suggested to be maximally efficient in radio-quiet clusters. Instead, about one half of the objects in our sample also have a central radio halo, at odds with the expectation of this model.

(vi) All models discussed in this paper assume a similar dependence between the acceleration efficiency of CR electrons and protons and the Mach number. However, recent studies using particle-in-cell simulations suggested the possibility of an efficient acceleration of electrons in quasi-perpendicular shocks by oblique whistler waves excited in the shock foot \citep[][]{2011ApJ...733...63R}. Moreover, it has been shown that the acceleration of electrons and ions might require different magnetic orientations and that the two species can be accelerated in different regions of the shock\citep[][]{2012ApJ...744...67G}. In order to reconcile the observed relic synchrotron emission and the non-detection of hadronic emission, an assumed oblique would magnetic field at shocks would have to lead to an efficient acceleration of electrons but not of protons. However, owing to the present theoretical and numerical difficulties in investigating the process of particle acceleration at the ''microphysical`` scales, little can be said about the validity of this scenario in galaxy clusters.\\

Certainly, future deeper limits from the FERMI satellite or other $\gamma$-ray telescopes will greatly enhance
the significance of the limits we derived here. 
%While the modelling of single objects can often be prone to a combination of "ad-hoc" solution, a comparison with a large number of objects provides a more robust test for the theory. 

\section*{acknowledgments}

We gratefully acknowledge B. Huber and D. Eckert for making their $\gamma$-ray upper limits available to us before the publication of their paper.
FV and MB acknowledge support from the grant FOR1254 from the Deutsche Forschungsgemeinschaft. MB acknowledges allocations 5056 and 5984 on supercomputers
at the NIC of the Forschungszentrum J\"{u}lich. 
We thank our referee for the useful comments that improved the quality of this paper, and  H. Kang, T. Jones, G. Brunetti, A. Bonafede \& J. Donnert  for their helpful feedback. We are indebted with M. Trasatti for pointing out a few typos in the Equations written in the first version of the paper.

\bibliographystyle{mnras}
\bibliography{franco}

\end{document}